\documentclass[conference]{IEEEtran}
\IEEEoverridecommandlockouts
\usepackage[T1]{fontenc} %
\usepackage{amsmath}
\usepackage[cmintegrals]{newtxmath}
\usepackage{bm} %
\usepackage{url}
\usepackage{mdwtab}
\usepackage{siunitx}
\makeatletter
\let\MYcaption\@makecaption
\makeatother
\usepackage[font=footnotesize]{subcaption}

\makeatletter
\let\@makecaption\MYcaption
\makeatother

\DeclareMathOperator{\E}{\mathbb{E}}
\usepackage[acronym,nopostdot,style=index,nonumberlist,toc]{glossaries}
\usepackage{graphicx}

\usepackage{amsmath}

\usepackage{svg}
\usepackage{mathtools}

\usepackage{breqn}   %

\usepackage{booktabs}
\usepackage{tabularx}
\newacronym{mmwave}{mmWave}{millimeter wave}
\newacronym{ai}{AI}{artificial intelligence}
\newacronym{rrm}{RRM}{radio resource management}
\newacronym{embb}{eMBB}{enhanced mobile broadband}
\newacronym{urllc}{uRLLC}{ultra-reliable and low-latency communications}
\newacronym{mmtc}{mMTC}{massive machine-type communications}
\newacronym{nr}{NR}{new radio}
\newacronym{5g}{5G}{fifth generation}
\newacronym{3gpp}{3GPP}{Third Generation Partnership Project}
\newacronym{bs}{BS}{base station}
\newacronym{ue}{UE}{user equipment}
\newacronym{dci}{DCI}{downlink control information}
\newacronym{uci}{UCI}{uplink control information}
\newacronym{dcm}{DCM}{downlink control message}
\newacronym{ucm}{UCM}{uplink control message}
\newacronym{pdcch}{PDCCH}{physical downlink control channel}
\newacronym{pdsch}{PDSCH}{physical downlink shared channel}
\newacronym{pucch}{PUCCH}{physical uplink control channel}
\newacronym{pusch}{PUSCH}{physical uplink shared channel}
\newacronym{dlsch}{DL-SCH}{downlink shared channel}
\newacronym{bch}{BCH}{broadcast channel}
\newacronym{pch}{PCH}{paging channel}
\newacronym{ul}{UL}{uplink}
\newacronym{dl}{DL}{downlink}
\newacronym{ulsch}{UL-SCH}{uplink shared channel}
\newacronym{rach}{RACH}{random-access channel}
\newacronym{mac}{MAC}{medium access control}
\newacronym{tdma}{TDMA}{time division multiple access}
\newacronym{tti}{TTI}{transmission time interval}
\newacronym{phy}{PHY}{physical layer}
\newacronym{bler}{BLER}{block error rate}
\newacronym{tb}{TB}{transport block}
\newacronym{tbs}{TBS}{transport block size}
\newacronym{tbler}{TBLER}{transport block error rate}
\newacronym{prb}{PRB}{physical resource block}
\newacronym{re}{RE}{resource element}
\newacronym{rb}{RB}{resource block}
\newacronym{sr}{SR}{scheduling request}
\newacronym{sg}{SG}{scheduling grant}
\newacronym{ack}{ACK}{acknowledgement}
\newacronym{rl}{RL}{reinforcement learning}
\newacronym{ml}{ML}{machine learning}
\newacronym{mdp}{MDP}{Markov decision process}
\newacronym{ql-amc}{QL-AMC}{Q-learning based adaptive modulation and coding}
\newacronym{ql-la}{QL-LA}{Q-learning based link adaptation}
\newacronym{dqn}{DQN}{deep Q-network}
\newacronym{ma}{MA}{multi-agent}
\newacronym{marl}{MARL}{multi-agent reinforcement learning}
\newacronym{mas}{MAS}{multi-agent systems}
\newacronym{maddpg}{MADDPG}{multi-agent deep deterministic policy gradient}
\newacronym{ddpg}{DDPG}{deep deterministic policy gradient}
\newacronym{ctde}{CTDE}{centralized training and decentralized execution}
\newacronym{sdu}{SDU}{service data unit}
\newacronym{pdu}{PDU}{protocol data unit}
\newacronym{ci}{CI}{confidence interval}
\newacronym{mlp}{MLP}{multilayer perceptron}
\newacronym{relu}{ReLU}{rectified linear unit}
\newacronym{dec-pomdp}{Dec-POMDP}{decentralized partially observable Markov decision process}

\newglossary[nlg]{notation}{not}{ntn}{Notation}

\newglossaryentry{not:nUE}{
   name=\ensuremath{L},
   description={number of users},
   type=notation
   }

\newglossaryentry{not:buffer-size}{
   name=\ensuremath{B},
   description={buffer size},
   type=notation
   }

\newglossaryentry{not:policy-net}{
   name=\ensuremath{\mu},
   description={policy network},
   type=notation
   }

\newglossaryentry{not:policy-params}{
   name=\ensuremath{\theta},
   description={policy network},
   type=notation
   }

\newglossaryentry{not:total-packets}{
   name=\ensuremath{P},
   description={number of packets to transmit},
   type=notation
   }

\newglossaryentry{not:reward-signal}{
   name=\ensuremath{\rho},
   description={number of packets to transmit},
   type=notation
   }

\newglossaryentry{not:p-arrival}{
   name=\ensuremath{p_{\mathrm{a}}},
   description={probability of arriving a new packet},
   type=notation
   }

\newglossaryentry{not:episode-duration}{
   name=\ensuremath{t_{\mathrm{max}}},
   description={maximum number of TTIs},
   type=notation
   }

\newglossaryentry{not:time-step}{
   name=\ensuremath{t},
   description={time step},
   type=notation
   }

\newglossaryentry{not:dl-vocabulary-size}{
   name=\ensuremath{D},
   description={downlink control message vocabulary size},
   type=notation
   }

\newglossaryentry{not:ul-vocabulary-size}{
   name=\ensuremath{U},
   description={uplink control message vocabulary size},
   type=notation
   }

\newglossaryentry{not:dl-message}{
   name=\ensuremath{m},
   description={downlink control message},
   type=notation
   }

\newglossaryentry{not:ul-message}{
   name=\ensuremath{n},
   description={uplink control message},
   type=notation
   }

\newglossaryentry{not:ue-action}{
   name=\ensuremath{a},
   description={UE action},
   type=notation
   }

\newglossaryentry{not:obs}{
  name=\ensuremath{o},
  description={observation},
  type=notation
  }

\newglossaryentry{not:state}{
  name=\ensuremath{x},
  description={agent state},
  type=notation
  }

\newglossaryentry{not:history-len}{
  name=\ensuremath{K},
  description={length of history buffer},
  type=notation
  }

\newglossaryentry{not:episodes-train}{
  name=\ensuremath{N_\mathrm{train}},
  description={number of training episodes},
  type=notation
  }

\newglossaryentry{not:episodes-eval}{
  name=\ensuremath{N_\mathrm{eval}},
  description={number of evaluation episodes},
  type=notation
  }

\newglossaryentry{not:episodes-test}{
  name=\ensuremath{N_\mathrm{test}},
  description={number of test episodes},
  type=notation
  }

\newglossaryentry{not:repetitions}{
  name=\ensuremath{N_\mathrm{rep}},
  description={number of test episodes},
  type=notation
  }

\newglossaryentry{not:n-receptions}{
  name=\ensuremath{N_\mathrm{SDUs}},
  description={number of test episodes},
  type=notation
  }

\newcommand{\nonl}{\renewcommand{\nl}{\let\nl\oldnl}}
\let\oldnl\nl%

\begin{document}

\title{The Emergence of Wireless MAC Protocols with Multi-Agent Reinforcement Learning

\thanks{Jakob Hoydis contributed to this work as a member of Nokia Bell Labs France until May 2021}
}

\author{
  \IEEEauthorblockN{
    Mateus P. Mota\IEEEauthorrefmark{1}\IEEEauthorrefmark{2},
    Alvaro Valcarce\IEEEauthorrefmark{1},
    Jean-Marie Gorce\IEEEauthorrefmark{2},
    Jakob Hoydis\IEEEauthorrefmark{3}
    }
  \IEEEauthorblockA{
    \IEEEauthorrefmark{1}Nokia Bell Labs, Nozay, France\\
    Email: mateus.pontes_mota@nokia.com, alvaro.valcarce_rial@nokia-bell-labs.com
   }
    \IEEEauthorblockA{
    \IEEEauthorrefmark{2}National Institute of Applied Sciences, Lyon, France\\
    Email: jean-marie.gorce@insa-lyon.fr
    }
    \IEEEauthorblockA{
    \IEEEauthorrefmark{3}NVIDIA, Paris, France\\
    Email: jhoydis@nvidia.com
    }

}
\maketitle

\begin{abstract}

In this paper, we propose a new framework, exploiting the \gls{maddpg} algorithm, to enable a \gls{bs} and \gls{ue} to come up with a medium access control (MAC) protocol in a multiple access scenario.
In this framework, the \gls{bs} and \glspl{ue} are \gls{rl} agents that need to learn to cooperate in order to deliver data.
The network nodes can exchange control messages to collaborate and deliver data across the network, but without any prior agreement on the meaning of the control messages.
In such a framework, the agents have to learn not only the channel access policy, but also the signaling policy.
The collaboration between agents is shown to be important, by comparing the proposed algorithm to ablated versions where either the communication between agents or the central critic is removed.
The comparison with a contention-free baseline shows that our framework achieves a superior performance in terms of goodput and can effectively be used to learn a new protocol.

\end{abstract}


\glsresetall

\section{Introduction}
\label{sec:Intro}

The current 5G networks are designed to support a wide range of services, including \gls{embb}, \gls{urllc} and \gls{mmtc}, which in turn will support an important growth in the number of applications.
This upsurge in novel services and applications is expected to also happen with 6G \cite{6g_vision}.
This heterogeneity of wireless networks may represent a challenge to protocol design.
Therefore, protocols tailored to specific applications may perform better than general-purpose solutions \cite{pasandi2021}.

\Gls{ml} can be used to design protocols, boost the network capacity \cite{hoydis20216g} and reduce the efforts and costs for future standardization \cite{6g_standardization}.
It is possible to view a protocol as the language of a network, since the network nodes have to negotiate how to transmit data by exchanging messages.
Then, the idea of emerging a new protocol would be similar to emerging communication between the network nodes.
\Gls{rl} is one category of \gls{ml} that provides the means to reach this goal.

During the last years, research on how to emerge communication in order to achieve collaboration between multiple agents received a growing attention \cite{lazaridou2020emergent}.
This growth partly relies on recent advances in \gls{marl} for cooperative problems \cite{dafoe2020open}.
Learning to cooperate by leveraging communication is about teaching agents to either learn existing natural languages or to emerge a fully new communication protocol that would help them collaborate to solve a task.

\textbf{Contribution:}
Our proposal is to leverage cooperative \gls{marl} augmented with communication to allow a fully new \gls{mac} protocol to emerge.
The idea of learning a given protocol has already been addressed in a previous work \cite{alvaro2021}, but to the best of the authors' knowledge, there is no previous work on studying the emergence of a new \gls{mac} protocol (signalling included) with \gls{marl}.
In the future, this idea may be used to develop application-tailored protocols that could perform better than the human-designed ones.

This work is structured as follows.
In Section~\ref{sec:rel-work}, we briefly review the literature.
In Section~\ref{sec:background}, we give a short background overview of \gls{marl} and the algorithm used in this work.
Section~\ref{sec:system-model} describes the system model used and in Section~\ref{sec:proposed}, we present a new framework allowing the emergence of \gls{mac} protocols with \gls{marl}.
Finally, Section~\ref{sec:simulation} illustrates the performance of our algorithm with our numerical results, where we compare the proposed solution with a baseline.
The main conclusions are drawn in Section~\ref{sec:conclusion}.

\section{Related Work}
\label{sec:rel-work}

Several papers have applied \gls{rl} to the \gls{mac} layer, mostly to solve \gls{rrm} problems such as scheduling (\cite{leasch, rrs_deep_pointer}) and  dynamic spectrum access (\cite{rl_for_spectrum_access, wong2020dynamic}).

In \cite{alvaro2021}, \gls{marl} is used to learn a predefined protocol and a new channel access policy.
This is done by having a \gls{bs} which uses a predefined protocol while the \glspl{ue} are \gls{rl} agents.
The \glspl{ue} are trained to learn the signaling and how to access the channel without any prior knowledge.
This way, they can learn their own channel-access policy, while respecting the target signaling policy.
However, in this case the agents only learn to use an already known \gls{mac} signaling, rather than developing a new one.

In \cite{pasandi2021}, a framework to design a protocol is proposed by considering the different functions a \gls{mac} protocol must perform.
An \gls{rl} agent designs a protocol by selecting which building function to use according to the network conditions.
However, in this case, the \gls{rl} agent still has a prior knowledge due to the use of the predefined protocol functions.

In \cite{learning_phy, remote_rl}, cooperative \gls{marl} is used to emerge a coding scheme by joint learning of communication and cooperation to solve a task with the help of a noisy communication channel.
The proposition of both works is to emerge a coding scheme that is tailored to the application.
None of these works address the question of learning a new signaling protocol.

\section{Background on MARL}
\label{sec:background}

\Gls{rl} is an area of \gls{ml} that aims to find the best behavior for an agent interacting with a dynamic environment in order to maximize a notion of accumulated reward \cite{Bishop07}.
The goal of the \gls{rl} agent is to find the best policy, which is the mapping of the perceived states to the actions to be taken.
The action-value function $Q^{\pi}(s_t,a_t)$, also known as Q-function, is the overall expected reward for taking action $a_t$ in state $s_t$ and then following a policy $\pi$.

\begin{figure}[tb]
    \vspace*{3pt}
    \centering
    \includegraphics[width=\columnwidth]{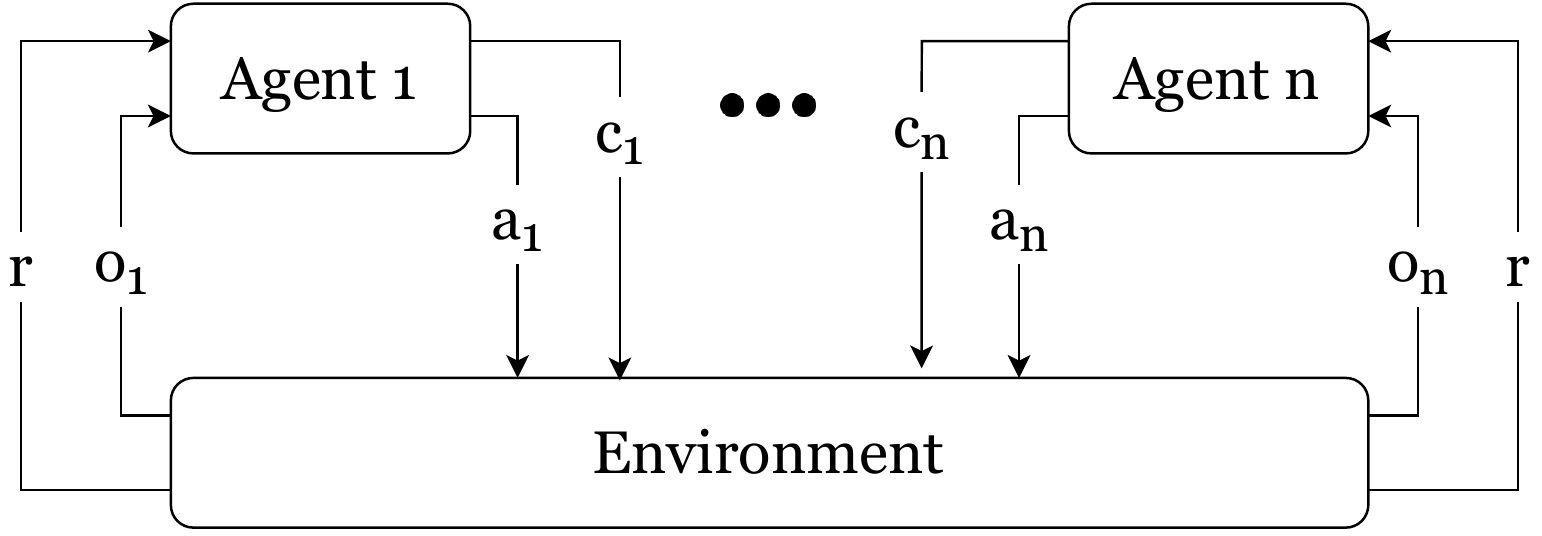}
    \caption{Cooperative MA-RL scheme with communication,}
    \label{fig:marl-scheme}
\end{figure}

MARL is an extension of RL for to \gls{mas}, where multiple agents interact with a system, i.e the environment.
In this work, we use the \gls{dec-pomdp} formulation \cite{oliehoek2008optimal}, augmented with communication.
A \gls{dec-pomdp} for $n$ agents is defined by the global state space $\mathcal{S}$, an action space $\mathcal{A}_1 , \ldots , \mathcal{A}_n $, and an observation space  $\mathcal{O}_1 , \ldots , \mathcal{O}_n $ for each agent.
In \gls{dec-pomdp}, the agent observation does not fully describe the environment state.
All agents share the same reward and the action space of each agent is subdivided into one environment action space and a communication action space.
The communication action represents the message sent by an agent and it does not affect the environment directly, but it may be passed to other agents.
This formulation is shown in Fig.~\ref{fig:marl-scheme}, where $o_i$ represents the observation received by the $i^{\mathrm{th}}$ agent, $r$ represents the reward, $a_i$ and $c_i$ represent the environment and communication actions, respectively.
In this work, the agent internal state $x_i$ may comprise not only the agent's current observation, but also previous observations, actions and received messages.

\Gls{marl} introduces some new challenges, such as partial observability and non-stationarity \cite{MARL_Challenges}.
In this work, we adopt the \gls{maddpg} algorithm \cite{lowe2017multi}, an extension of the \gls{ddpg} algorithm \cite{ddpg} to multi-agent problems with \gls{ctde}.
It addresses the non-stationarity problem by using a centralized critic.
Each agent has an actor network that depends only on its own agent's state in order to learn a decentralized policy $\mu_i$ with parameters $\theta_i$.
During the training, each agent has a centralized critic that receives the agent states and actions of all agents in order to learn a joint action value function $Q_i (x,a)$ with parameters $\varphi_i$, where $x = (x_1, x_2, \ldots, x_n)$ is a vector containing all the agents' states and $a = (a_1, a_2, \ldots, a_n)$ contains the actions taken by all of the agents.

The critic network parameters $\varphi$ are updated by minimizing the loss given by the temporal-difference error
\begin{equation}
\label{eq.:critic}
    L^i \coloneqq \E_{x, a, r, x^{\prime} \thicksim \mathcal{D} } \left[ y^i - Q_i (x, a_1, \ldots, a_n ; \varphi_i)  \right]
\end{equation}
\noindent where $\mathcal{D}$ denotes the experience replay buffer in which the transition tuples $ (x, a, r, x^{\prime})$ are stored, $Q^{\prime}$ and $\mu^{\prime}$ represent the target critic network and the value of the target actor network, with parameters $\theta^{\prime}$ and $\varphi^{\prime}$, respectively, and $y^i$ is the temporal-difference target, given by
\begin{equation}
\label{eq.:td-error}
    y^i \coloneqq r + \left. \gamma Q^{\prime}_i ( x^{\prime}, a^{\prime}_1,  \ldots, a^{\prime}_n; \varphi^{\prime}_i )  \right|_{a^{\prime}_k = \mu^{\prime}_k (x_k)}
\end{equation}
\noindent where $\gamma$ is the discount factor.
The actor network parameters $\theta$ are updated using the sampled policy gradient
\begin{equation}
\label{eq.:actor}
    \nabla_{\theta_i} J = \E_{x, a \thicksim \mathcal{D} } \left[ \nabla_{a_i} Q_i (x,a) \nabla_{\theta_i} \mu_i (x_i) \mid a_i=\mu_i (x_i) \right] .
\end{equation}
The target networks parameters are updated as
\begin{equation}
\label{eq.:target-critic}
    \varphi^{\prime}_i  \leftarrow \tau \varphi_i + (1-\tau) \varphi^{\prime}_i
\end{equation}
\begin{equation}
\label{eq.:target-actor}
    \theta^{\prime}_i  \leftarrow \tau \theta_i + (1-\tau) \theta^{\prime}_i
\end{equation}
\noindent where $\tau \in \left[ 0 , 1 \right]$ is the soft-update parameter.
Smaller values of $\tau$ lead to slow target network changes and are generally preferred \cite{ddpg}.

\section{System Model}
\label{sec:system-model}

We consider a single cell with a \gls{bs} serving \gls{not:nUE} \glspl{ue} operating according to a \gls{tdma} scheme, where each \gls{ue} needs to deliver \gls{not:total-packets} \glspl{sdu} to the \gls{bs}.
We assume that each \gls{mac} \gls{pdu} contains only one \gls{sdu}.
The network nodes can communicate, i.e. exchange information, using messages through the control channels.
In the rest of this paper, we use the expressions \gls{ue} and \gls{bs} to refer to the \gls{ue} \gls{mac} agent and the \gls{bs} \gls{mac} agent, respectively.

The channel for the uplink data transmission is modeled as a packet erasure channel, where a \gls{tb} is incorrectly received with a probability given by a \gls{tbler}.
The \glspl{dcm} and \glspl{ucm} are transmitted over the \gls{dl} and \gls{ul} control channels, which are assumed to be error free and without any contention or collision.

\begin{figure}[tb]
    \vspace*{3pt}
    \centering
    \includegraphics[width=0.98\columnwidth]{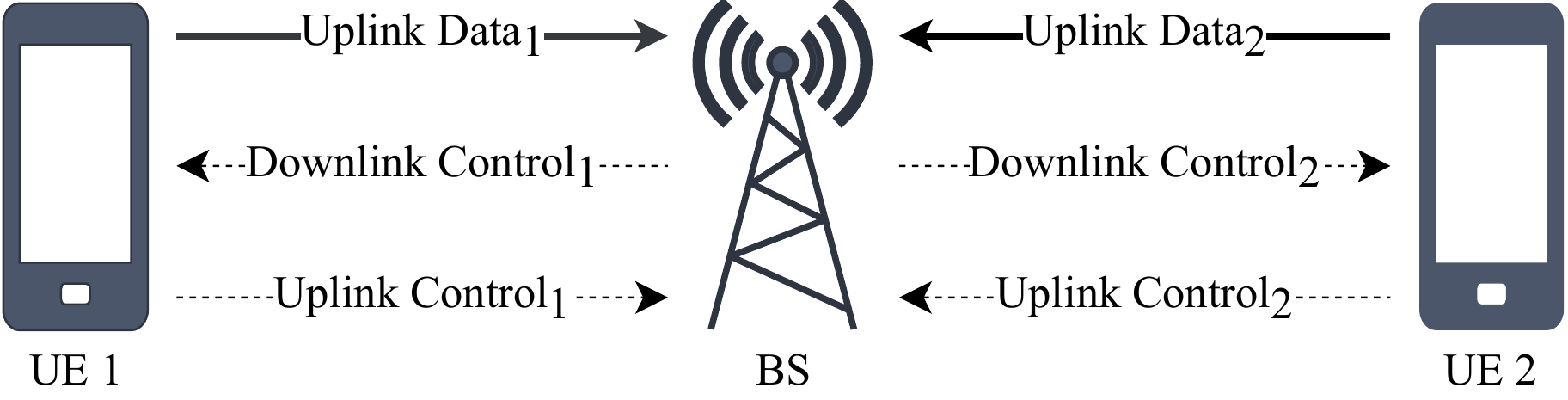}
    \caption{System model example with two \glspl{ue}.}
    \label{fig:sys-model}
\end{figure}

We assume that the sets of possible \gls{dl} and \gls{ul} control messages have cardinality \gls{not:dl-vocabulary-size} and \gls{not:ul-vocabulary-size}, respectively.
For example, the \glspl{ucm} in an \gls{ul} control vocabulary of size $\gls{not:dl-vocabulary-size} = 8$ would have a bitlength of $\log_2 {\gls{not:dl-vocabulary-size}}=3$.
This exchange is shown in Fig.~\ref{fig:sys-model}, where dashed lines indicate control information and solid lines indicate user data.

Each \gls{ue} has a transmission buffer of capacity \gls{not:buffer-size} \gls{mac} \glspl{sdu} that starts empty.
At each time step \gls{not:time-step}, a new \gls{sdu} is added to the buffer with probability \gls{not:p-arrival}, until a maximum of \gls{not:total-packets} \glspl{sdu} have been generated for each \gls{ue}.

At each time step \gls{not:time-step}, the \gls{bs} can send one control message to each \gls{ue} and each \gls{ue} can send one control message to the \gls{bs} while being able to send data \glspl{pdu} through the \gls{ulsch}.
Furthermore, the \glspl{ue} can also delete a \gls{sdu} from the buffer at each time step.

The transmission task is considered finished once all \glspl{sdu} are received and all transmission buffers are empty.
We define the goodput $G$ (in \glspl{sdu}/TTIs) as the number of \gls{mac} \glspl{sdu} received by the \gls{bs} per unit of time, without considering the retransmissions:
\begin{equation}
    G = \frac{N_{\mathrm{RX}}} {N_{\mathrm{TTIs}}}
\end{equation}
\noindent where $N_{\mathrm{RX}}$ represents the number of \glspl{sdu} received and $N_{\mathrm{TTIs}}$ is the total time taken to finish the transmission task.
The delivery-rate $\Gamma_{\mathrm{RX}}$ is the percentage of \glspl{sdu} correctly received by the \gls{bs}:
\begin{equation}
    \label{eq:delivery-rate}
    \Gamma = \frac{N_{\mathrm{TX}}}{\gls{not:total-packets} \gls{not:nUE}} \text{.}
\end{equation}

\section{Emerging a MAC Protocol with MARL}
\label{sec:proposed}

\subsection{MARL Formulation}

We can formulate the problem defined above as a \gls{marl} cooperative task, where the \gls{mac} layers of the network nodes (\glspl{ue} and \gls{bs}) are \gls{rl} agents that need to learn how to communicate with each other to solve an uplink transmission task.
In addition, the \gls{ue} agents need to learn when to send data through the \gls{ulsch} and when to delete an \gls{sdu}, in other words, to learn how to correctly manage the buffer.

In order to decide how to act, an agent needs to consider the messages received from the other agents. %
In addition, the \glspl{ue} also take into account their buffer status when taking actions, while the \gls{bs} takes into account the state of the \gls{ulsch}, i.e idle, busy or collision-free reception.

We use the following notations:
\begin{itemize}

    \item $\gls{not:obs}^{\mathrm{u}}_{t}$: Observation received by the $u$\textsuperscript{th} \gls{ue} at time step \gls{not:time-step}.
    \item $\gls{not:obs}^{\mathrm{b}}_{t}$: Observation received by the \gls{bs} at time step \gls{not:time-step}.
    \item $\gls{not:ul-message}^{\mathrm{u}}_{t}$: The \gls{ucm} sent from the $u^{\mathrm{th}}$ \gls{ue} at time step $t$.
    \item $\gls{not:dl-message}^{\mathrm{u}}_{t}$: The \gls{dcm} sent to the $u^{\mathrm{th}}$ \gls{ue} at time step $t$.
    \item $\gls{not:ue-action}^{\mathrm{u}}_{t}$: Environment action of the $u^{\mathrm{th}}$ \gls{ue} at time step $t$.
    \item $\gls{not:state}^{\mathrm{u}}_{t}$: Agent internal state of the $u^{\mathrm{th}}$ \gls{ue} at time step $t$.
    \item $\gls{not:state}^{\mathrm{b}}_{t}$: Agent internal state of the \gls{bs} at time step $t$.

\end{itemize}

The observation $\gls{not:obs}^{\mathrm{u}}_{t} \in \left\{ 0, \dots, \gls{not:buffer-size} \right\} $ is a integer representing the number of \glspl{sdu} in the buffer of the \gls{ue} $u$ at that time $t$.
Similarly, the observation $\gls{not:obs}^{\mathrm{b}}_{t} $ received by the \gls{bs} is a discrete variable with $\gls{not:nUE} + 2$ possible states:
\begin{equation}\label{eq.:base-obs}
    \gls{not:obs}^{\mathrm{b}}_{t} = \begin{cases}
    0, \text{ if the \gls{ulsch} is idle} \\
    \mathrm{u}, \text{ } \parbox[t]{.33\textwidth}{ if the \gls{ulsch} is detected busy with a single \gls{pdu} from \gls{ue} $\mathrm{u}$, correctly decoded} \\
    \gls{not:nUE} + 1, \text{ non-decodable energy in the \gls{ulsch}} \\
    \end{cases}
\end{equation}
\noindent where $\mathrm{u} \in \left\{ 0, \dots, \gls{not:nUE} \right\}$.
The environment action $\gls{not:ue-action}^{\mathrm{u}}_{t}  \in \{0, 1, 2\}$ is interpreted as follows:
\begin{equation}
\label{eq.:action}
    \gls{not:ue-action}^{\mathrm{u}}_{t} = \begin{cases}
    0 \text{: do nothing} \\
    1 \text{: transmit the oldest \gls{sdu} in the buffer} \\
    2 \text{: delete the oldest \gls{sdu} in the buffer} \\
\end{cases}
\end{equation}

We assume the episode ends when all the \glspl{sdu} are correctly received by the \gls{bs} or when a maximum number of steps \gls{not:episode-duration} is reached.
The reward given at each time step is:
\begin{equation}
\label{eq.:reward}
    r_{t} = \begin{cases}
    +\gls{not:reward-signal}, \text{ if a new \gls{sdu} was received by the \gls{bs} } \\
    -\gls{not:reward-signal}, \text{ } \parbox[t]{.25\textwidth}{ if a \gls{ue} deleted a \gls{sdu} that has not been received by the \gls{bs}} \\
    -1, \text{ else,}
\end{cases}
\end{equation}
\noindent where \gls{not:reward-signal} is a positive integer.
This choice of reward is possible by leveraging the \gls{ctde}.
During the centralized training, a centralized reward system can be used to observe the buffers of the \gls{bs} and \glspl{ue} in order to assign the reward.

\subsection{Training Algorithm}
\label{sec:solution}

The proposed \gls{rl} solution is based on the \gls{maddpg} algorithm \cite{lowe2017multi}.
Each entity of the system has its own actor network which outputs the action of an agent given its state.
Each agent also has a centralized critic network which outputs the Q-value given the actions and states of all agents.
The critic networks are only used during the centralized training.

The actor and critic networks have the same architecture; a fully connected \gls{mlp} with two hidden layers, of $64$ neurons each.
The activation function of all hidden layers is the \gls{relu}.

The agent state at time step $t$ is a tuple comprising the most recent $k$ observations, actions and received messages:
\begin{itemize}
    \item \Gls{ue} $\mathrm{u}$: $\gls{not:state}^{\mathrm{u}}_{t} = ( \gls{not:obs}^{\mathrm{u}}_{t}, \ldots , \gls{not:obs}^{\mathrm{u}}_{t-k} , \gls{not:ue-action}^{\mathrm{u}}_{t}, \ldots, \gls{not:ue-action}^{\mathrm{u}}_{t-k}, \gls{not:ul-message}^{\mathrm{u}}_{t}, \ldots, \gls{not:ul-message}^{\mathrm{u}}_{t-k} , \\ \gls{not:dl-message}^{\mathrm{u}}_{t}, \ldots, \gls{not:dl-message}^{\mathrm{u}}_{t-k} ) $
    \item \Gls{bs}: $ \gls{not:state}^{\mathrm{b}}_{t} = ( \gls{not:obs}^{\mathrm{b}}_{t}, \ldots , \gls{not:obs}^{\mathrm{b}}_{t-k} , \mathbf{\gls{not:ul-message}}_{t}, \ldots, \mathbf{\gls{not:ul-message}}_{t-k} , \mathbf{\gls{not:dl-message}}_{t}, \ldots, \mathbf{\gls{not:dl-message}}_{t-k} ) $, with $ \mathbf{\gls{not:ul-message}}$ and $\mathbf{\gls{not:dl-message}}$ containing the messages from all the \glspl{ue}.
\end{itemize}

In order to improve training of our \gls{maddpg} solution, we make use of parameter sharing \cite{foerster2016learning} for similar network nodes, in this case the \glspl{ue}.
Similarly to the original work \cite{lowe2017multi}, we use the Gumbel-softmax \cite{gumbel-softmax} trick to soft-approximate the discrete actions to continuous ones.
The Gumbel-softmax reparameterization also works to balance exploration and exploitation.
The exploration-exploitation trade-off is controlled by the temperature factor $\zeta$.

After training finishes, we have successfully trained a population of $\gls{not:repetitions}=32$ protocols.
We then select the protocol that performed best during the last $\gls{not:episodes-eval}=500$ evaluation episodes.
This selection step can be seen as a "survival of the fittest" approach because only one protocol of the population of \gls{not:repetitions} is chosen going forward.

\section{Results}
\label{sec:simulation}
\subsection{Simulation Parameters}

\begin{figure*}[!tb]
    \centering
    \begin{subfigure}[t]{0.49\textwidth}
        \centering
        \includegraphics[width=\textwidth]{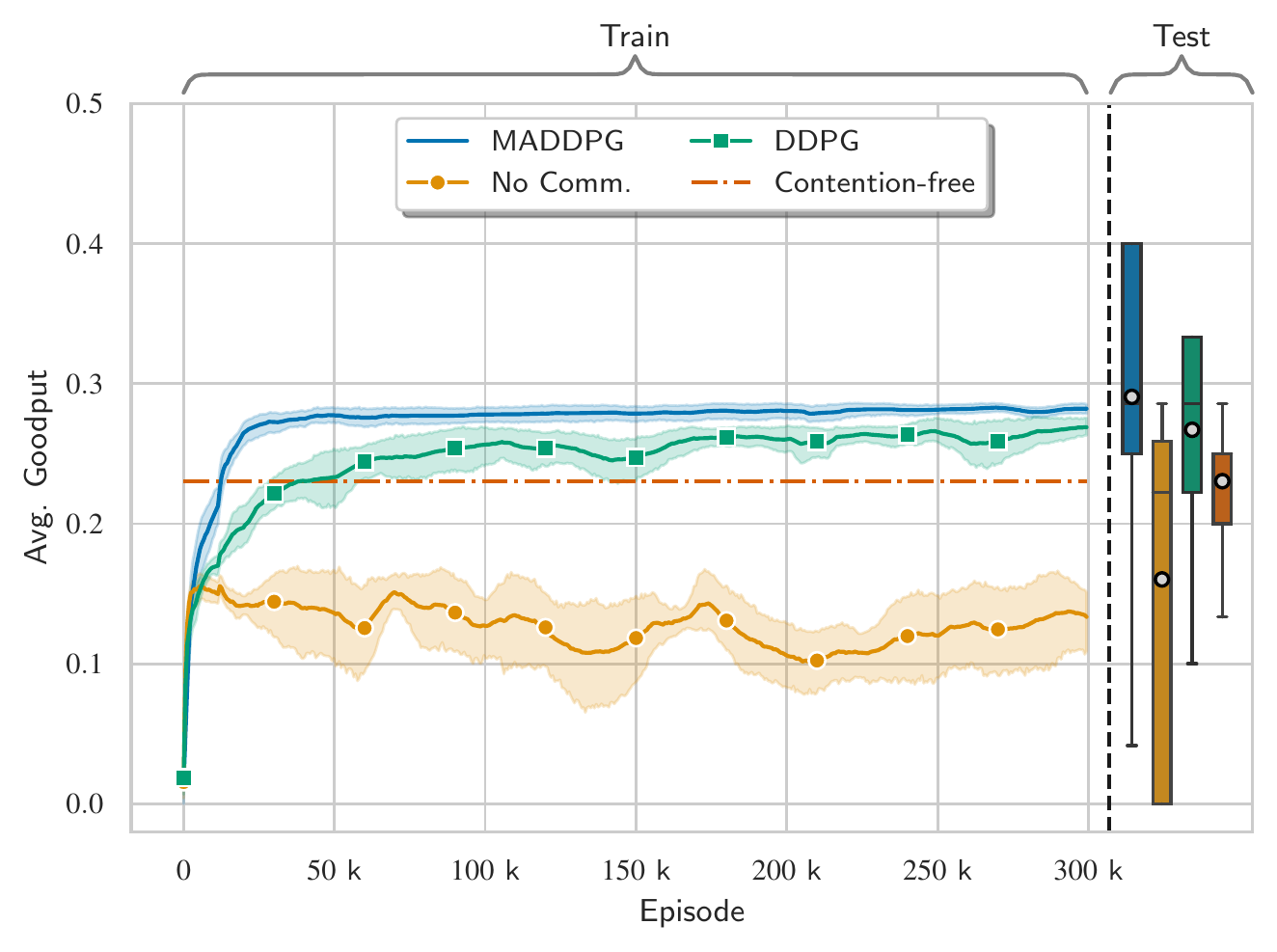}
        \caption{Number of SDUs: $\gls{not:total-packets}=1$}
        \label{fig:goodput-1-sdu}
    \end{subfigure}%
    ~
    \begin{subfigure}[t]{0.49\textwidth}
        \centering
        \includegraphics[width=\textwidth]{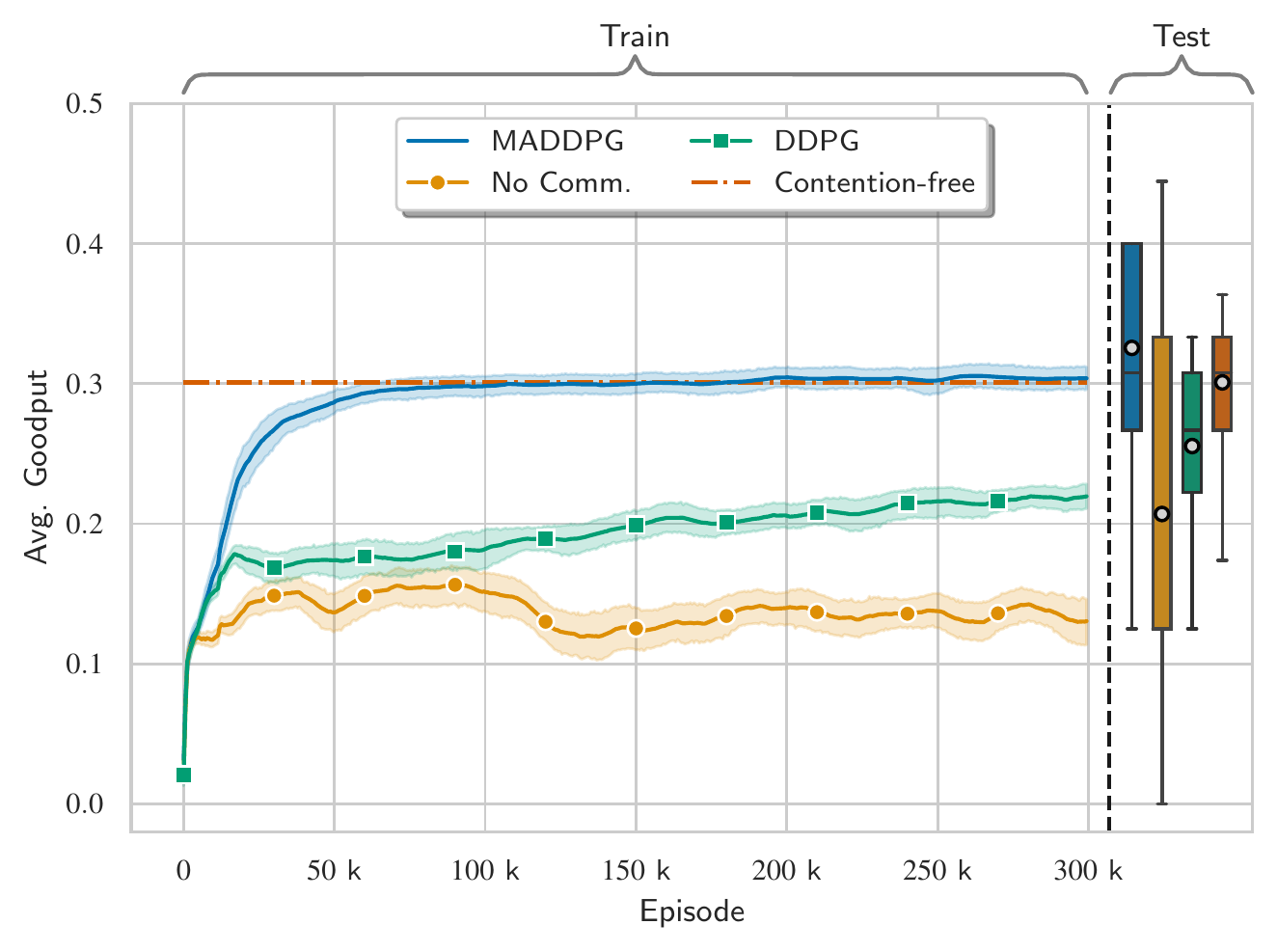}
        \caption{Number of SDUs: $\gls{not:total-packets}=2$}
        \label{fig:goodput-2-sdu}
    \end{subfigure}
    \caption{Goodput comparison.}
    \label{fig:train-results-goodput}
\end{figure*}

For simplicity, we assess the performance of a system with one \gls{bs} and two \glspl{ue}.
The transmission buffer of each user starts empty and the \gls{sdu} arrival probability \gls{not:p-arrival} is $0.5$.

The system is trained for a fixed number of episodes \gls{not:episodes-train}.
At some points during the training, we evaluate the policy on a total of \gls{not:episodes-eval} evaluation episodes with disabled exploration and disabled learning in order to assess the current performance of the \gls{mac} protocol.
The set of evaluation episodes remain the same in order to effectively compare the performance on the same set.
At the end of the training procedure, we further evaluate the learned protocol by assessing its performance in \gls{not:episodes-test} episodes with exploration and learning disabled.
This whole procedure represents a single training repetition.
We evaluate a total of \gls{not:repetitions} repetitions, each with a different random seed.

A summary of the main simulation parameters is provided in Table~\ref{tab:sim-params}, while the parameters of the \gls{maddpg} and \gls{ddpg} algorithms are listed in Table~\ref{tab:rl-params}.

\begin{table}[tb]
\centering
\caption{Simulation Parameters}
\label{tab:sim-params}
\begin{tabularx}{0.98\columnwidth}{l c c}
\toprule
\textbf{Parameter}                      & \textbf{Symbol} 	             & \textbf{Value}           \\
\midrule
Number of \glspl{ue}                    & \gls{not:nUE}                 &  $2$                      \\
Size of transmission buffer             & \gls{not:buffer-size}         &  $5$                      \\
Number of \glspl{sdu} to transmit       & \gls{not:total-packets}       &  $\left[ 1, 2 \right]$    \\
\Gls{sdu} arrival probability           & \gls{not:p-arrival}           &  $0.5$                    \\
Transport block error rate              & \gls{tbler}                   &  $\left[ 10^{-1}, 10^{-2}, 10^{-3}, 10^{-4} \right]$               \\
\gls{dcm} vocabulary size               & \gls{not:dl-vocabulary-size}  &  $3$                 \\
\gls{ucm} vocabulary size               & \gls{not:ul-vocabulary-size}  &  $2$                 \\
Max. duration of episode (\glspl{tti})  & \gls{not:episode-duration}    &  $24$                \\
Reward function parameter               & \gls{not:reward-signal}       &  $3$                 \\
Number of training episodes             & \gls{not:episodes-train}      &  $300 \si{k}$        \\
Number of evaluation episodes           & \gls{not:episodes-eval}       &  $500$               \\
Number of test episodes                 & \gls{not:episodes-test}       &  $5000$              \\
Number of randomized repetitions        & \gls{not:repetitions}         &  $32$                \\
\bottomrule
\end{tabularx}
\end{table}

\subsection{Baseline Solutions}

We compare the proposed solution with a contention-free baseline.
We also compare the proposed solution to two simplified approaches where either the communication between agents is not permitted or the centralized critic is disabled, i.e the \gls{ddpg} algorithm.
The ablation comparison helps to evaluate if communication and the centralized critic are needed to solve this task.

In the contention-free protocol, the \gls{ue} sends an \gls{sr} if its transmission buffer is not empty and it only transmits if it has received a \gls{sg}.
Similarly, it only deletes a \gls{tb} from the transmission buffer after the reception of an \gls{ack}.
At each time step, the \gls{bs} receives zero or more \glspl{sr}.
It then chooses one of the requesters at random to transmit in the next time-step, sending a \gls{sg} to the selected \gls{ue}.
However, if the \gls{ue} had made a successful data transmission simultaneously with an \gls{sr}, the \gls{bs} will send an \gls{ack} to this \gls{ue} and its \gls{sr} is ignored.

\begin{table}[tb]
\centering
\caption{Training Algorithm Parameters}
\label{tab:rl-params}
\begin{tabularx}{0.8\columnwidth}{X c}
\toprule
\textbf{Parameter}                      & \textbf{Value} 	            \\
\midrule
Memory length                           & 3                             \\
Replay buffer size                      & $10^5$                        \\
Batch size                              & 1024                          \\
Number of neurons per hidden layer      & $\{ 64, 64 \} $      \\
Interval between updating policies      & 96                            \\
Optimizer algorithm                     & Adam \cite{kingma2014adam}    \\
Learning rate                           & $10^{-3}$                     \\
Discount factor                         & $0.99$                        \\
Policy regularizing factor              & $10^{-3}$                     \\
Gumbel-softmax temperature factor       & $1$                           \\
Target networks soft-update factor      & $10^{-3}$                     \\

\bottomrule
\end{tabularx}
\end{table}

\subsection{Results}

\begin{figure*}[!tb]
    \centering
    \begin{subfigure}[t]{0.49\textwidth}
        \centering
        \includegraphics[width=\textwidth]{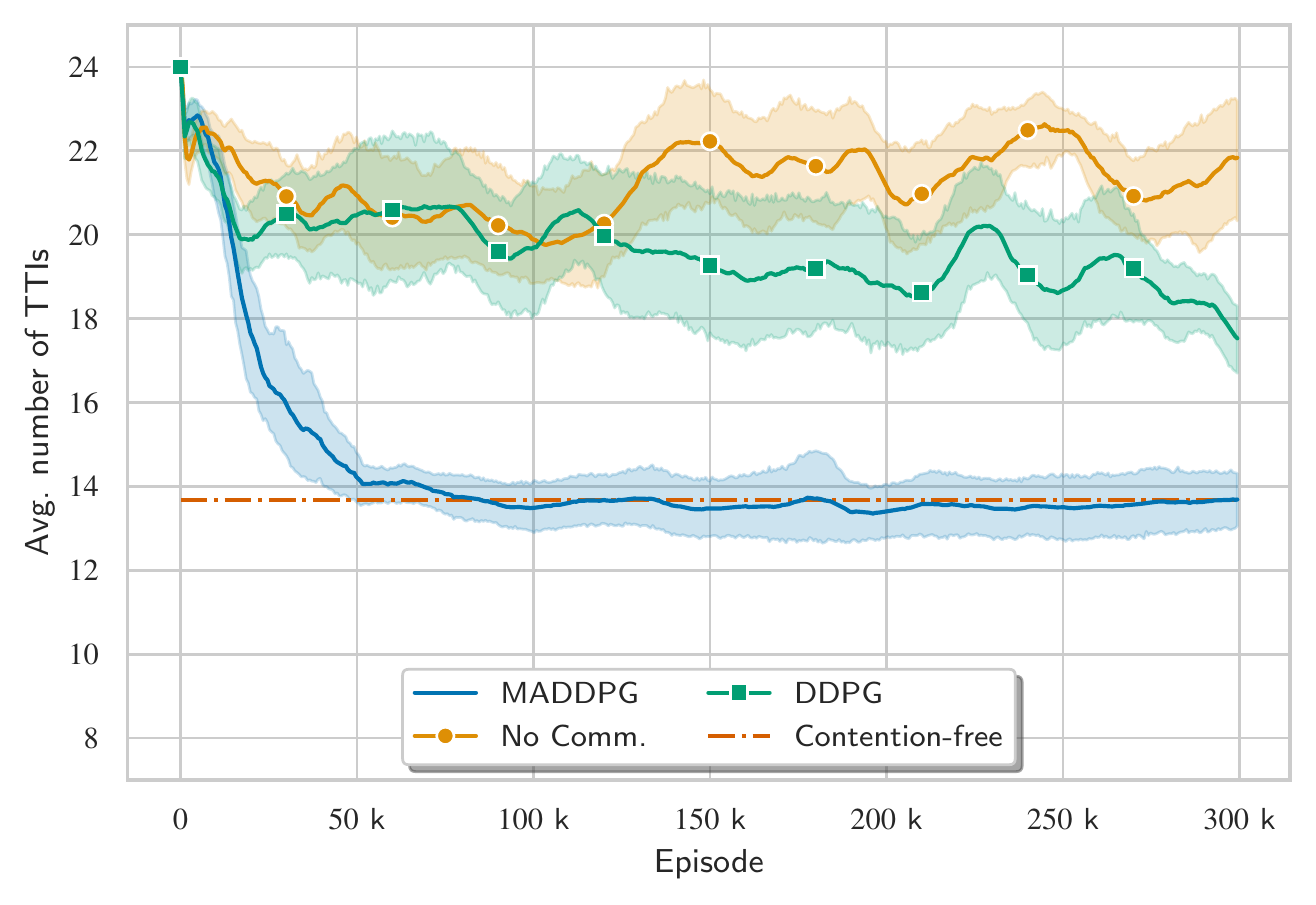}
        \caption{Episode duration in TTIs}
        \label{fig:episode-duration}
    \end{subfigure}%
    ~
    \begin{subfigure}[t]{0.49\textwidth}
        \centering
        \includegraphics[width=\textwidth]{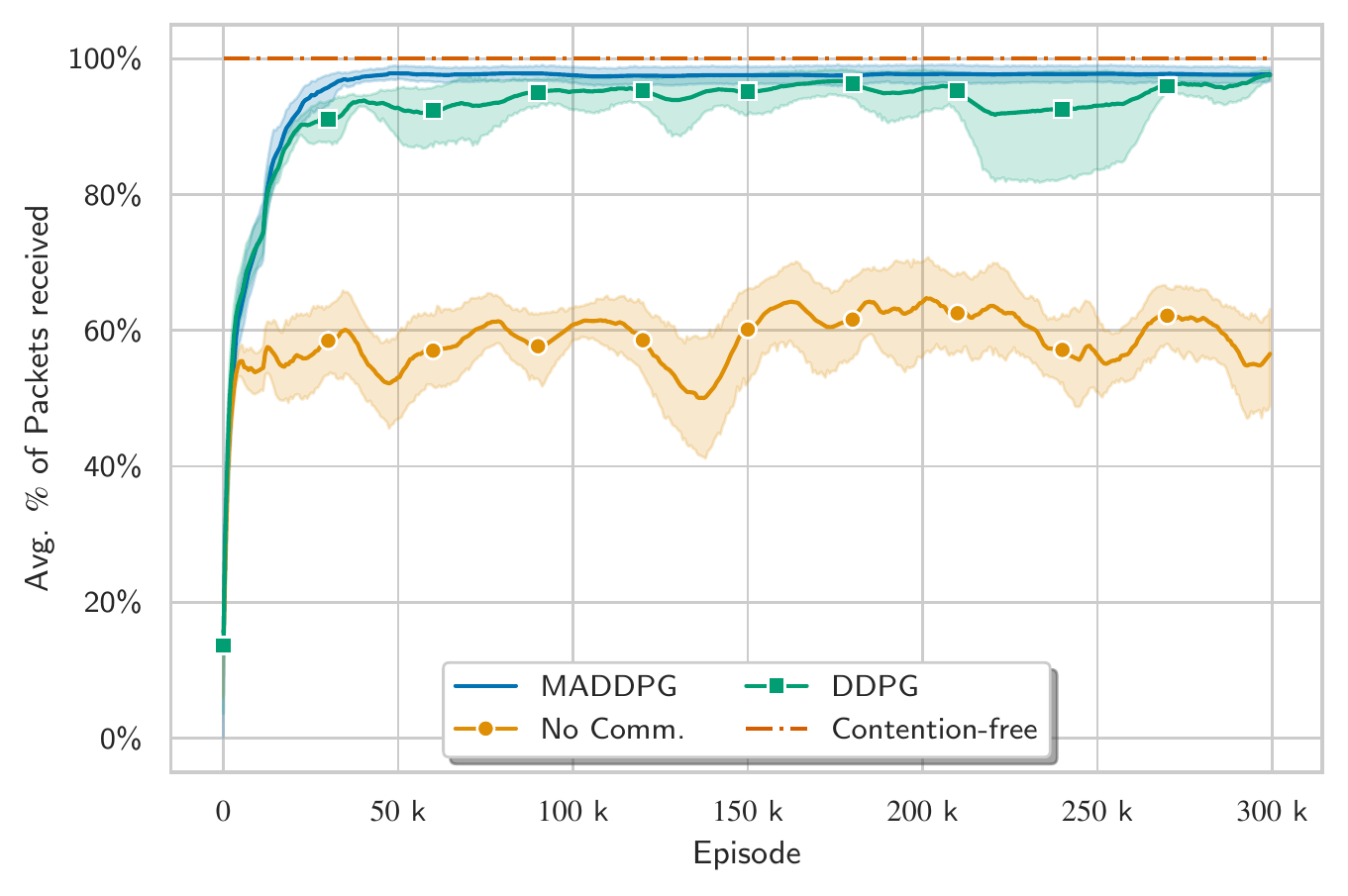}
        \caption{Pct. of Packets received}
        \label{fig:pct-packets}
    \end{subfigure}
    \caption{Performance comparison for two SDUs.}
    \label{fig:train-results-deep}
\end{figure*}

We compare the \gls{maddpg} solution with three other solutions, the contention-free baseline, the \gls{maddpg} solution without communication, and the \gls{ddpg} version of the proposed solution, i.e. the proposed solution without the centralized critic.
For the \gls{rl} solutions, the solid lines in Figs.~\ref{fig:train-results-goodput} and~\ref{fig:train-results-deep} show the average performance in the evaluation episodes during the training and the shaded areas represent the $95 \%$ \gls{ci}.
The dashed lines show the average performance of the baseline.

In Fig.~\ref{fig:train-results-goodput}, we compare the performance in terms of goodput for the \gls{tbler} of $10^{-1}$.
Figures~\ref{fig:goodput-1-sdu} and~\ref{fig:goodput-2-sdu} show the results when the \glspl{ue} have to transmit one and two \glspl{sdu}, respectively.
After assessing the performance on the last \gls{not:episodes-eval} evaluation episodes, we select the best performing repetitions for each solution in terms of average goodput to compare using boxplots of the test episodes.

By comparing the \gls{rl} solutions in both cases, the \gls{maddpg} has the best performance and the ablation without communication has the worst performance overall.
In addition, the \gls{maddpg} shows a more stable performance during training, with less variation than both other \gls{rl} solutions.
The ablation without communication has the greatest variation of performance, demonstrated by the \glspl{ci} and by the boxplots, indicating that communication helps achieving a more robust solution.

In Fig.~\ref{fig:goodput-1-sdu}, both the \gls{maddpg} and \gls{ddpg} solutions outperform the contention-free baseline, whereas the ablation without communication fails to effectively solve the task in this case.
When we move from one \gls{sdu} to two \glspl{sdu}, in Fig.~\ref{fig:goodput-2-sdu}, the \gls{ddpg} solution does not outperform the baseline.
Moreover, the difference in performance between the \gls{maddpg} and the baseline is reduced.

To better understand the goodput results of Fig.~\ref{fig:goodput-2-sdu}, Fig.~\ref{fig:train-results-deep} shows the performance in terms of episode duration, Fig.~\ref{fig:episode-duration}, and of percentage of the total \glspl{sdu} received during the episode as defined in Eq.~\eqref{eq:delivery-rate}, Fig.~\ref{fig:pct-packets}.

As shown in Fig.~\ref{fig:pct-packets}, the \gls{ddpg} algorithm achieves a high performance in terms of delivery-rate, but it takes more time to solve the task, thus the lower performance in terms of goodput when compared with the \gls{maddpg} and the baseline.
Comparing the \gls{maddpg} with the contention-free solution in Fig.~\ref{fig:episode-duration}, the proposed solution achieves a better goodput by finishing the task in less \glspl{tti}.
The proposed solution has a delivery-rate lower than the contention-free baseline, although it is also very close to $100 \%$.

By applying "survival of the fittest" to pick the best protocol in terms of goodput, the delivery-rate difference becomes even lower than shown in Fig~\ref{fig:pct-packets}.
The best protocol produced by the proposed solution has an average delivery-rate on the test episodes of $\Gamma_{\textsc{maddpg}} = 99.973 \% $, whereas the average of the contention-free baseline is of $\Gamma_{\mathrm{base}} = 99.998 \% $

\begin{figure}[tb]
    \centering
    \includegraphics[width=0.98\columnwidth]{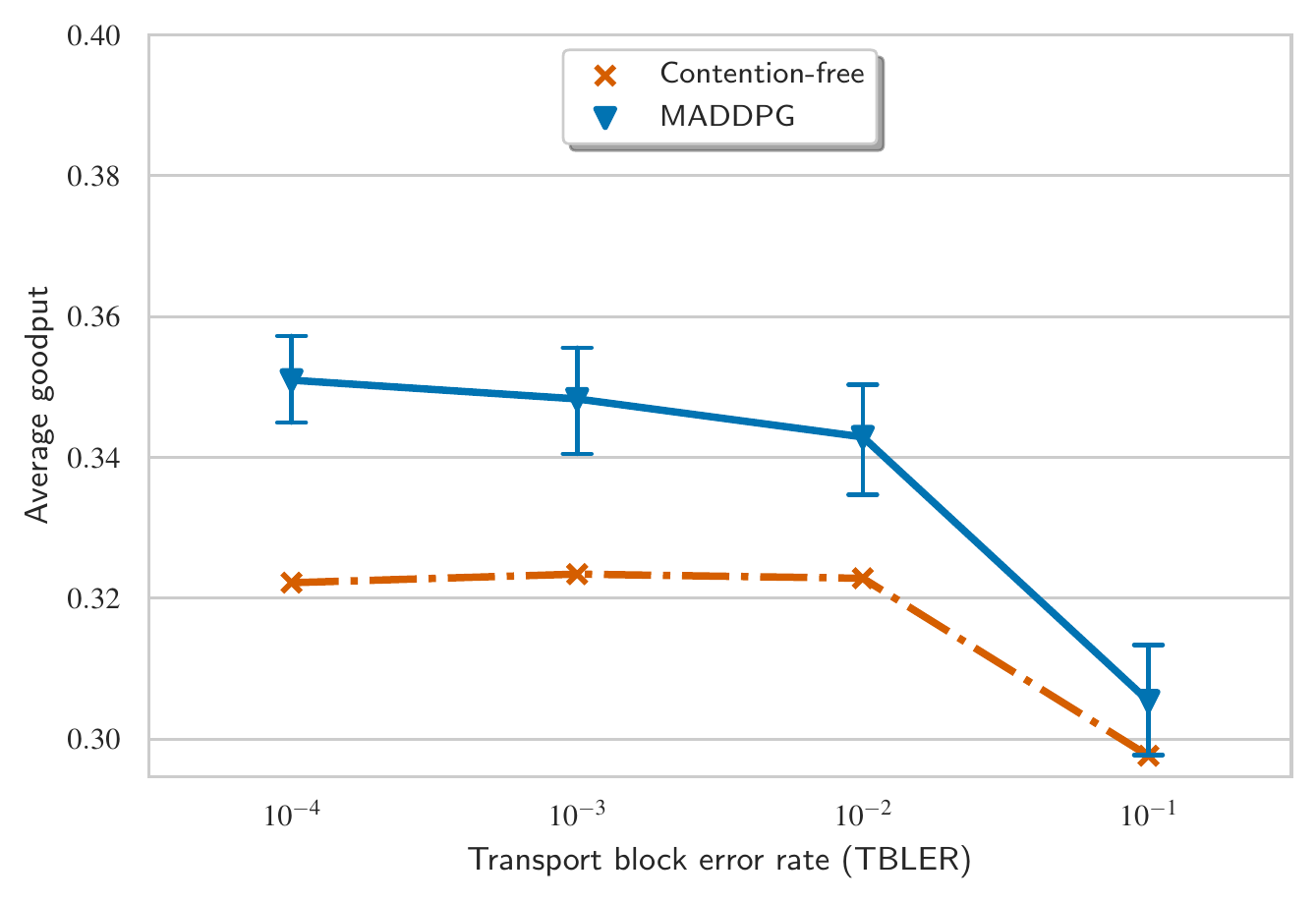}
    \caption{Performance in terms of goodput for different \glspl{tbler} with each \gls{ue} having to transmit two \glspl{sdu}.}
    \label{fig:goodput-bler}
\end{figure}

In Fig.~\ref{fig:goodput-bler}, we compare the proposed \gls{maddpg} framework with the contention-free baseline for different \glspl{tbler} and with each \gls{ue} having to transmit two \glspl{sdu}.
The performance is evaluated on \gls{not:episodes-test} test episodes by comparing the average goodput achieved.
For the \gls{maddpg} solution, we also show the $95\%$ \gls{ci} across randomized repetitions.
The proposed solution maintains a better performance than the baseline across the different \glspl{tbler}.
Moreover, the lowest difference in performance between the baseline and the proposed solution occurs when the \gls{tbler} is equal to $0.1$, showing that the proposed solution adapts well to lower \gls{tbler} regimes.

\section{Conclusions and Perspectives}
\label{sec:conclusion}

We have proposed a novel framework based on cooperative \gls{marl} augmented with communication, that provides us with the means to emerge a new protocol.
In essence, the agents have to learn the signaling policy, representing the control messages they exchange, and the channel-access policy, representing the \gls{phy} control of the agents.
Comparing with two ablations and a baseline, the results show that a solution capable to overcome the challenges in multi-agent systems is needed in order to emerge a protocol.
The results also indicate that enabling communication between agents is needed in order to solve a transmission task.
In addition, the results illustrate that the proposed solution can produce a protocol tailored to all \gls{tbler} regimes that outperforms a more general one.

Concerning future work, we highlight a study on the effect of the different parameters, such as the vocabulary sizes and \glspl{tbler}
Moreover, we envision a comparison with different \gls{marl} algorithms.
Finally, the application of this framework to a more complex system model is planned.

\section*{Acknowledgment}
\label{sec:ack}
\addcontentsline{toc}{section}{Acknowledgment}
The work of Mateus P. Mota is funded by Marie Sklodowska-Curie actions (MSCA-ITN-ETN 813999 WINDMILL).

\bibliographystyle{IEEEtran}
\bibliography{IEEEabrv, ref.bib}

\end{document}